\RequirePackage{fix-cm}
\documentclass[aps,pra,twocolumn,showpacs,preprintnumbers,amsmath,amssymb]{revtex4-1}
\usepackage{graphicx}
\usepackage{color,xcolor}
\usepackage{ulem}
\usepackage{amsmath}
\usepackage{subfig}

\definecolor{onera}{RGB}{21,150,215}

\input alphabet.tex   
\def\XS{\xspace}

\def\rem#1{}                    

\def\eg{\textit{e.g.,}\XS}

\def\+{^\dagger}        

\def\V{,\kern-.1em}     

\def\nequiv{\not\kern-.05em\equiv}
\def\egal{\kern-.5em=\kern-.5em}        

\def\reel#1{\mathrm{Re}}

\def\intdouble{\int\kern-0.65em\int}
\def\inttriple{\int\kern-0.65em\int\kern-0.65em\int}

\def\rond#1{\overset{\kern-0.33em~_\circ}{#1}}
\def\rondit[#1]#2{\overset{\kern#1~_\circ}{#2}}

\newsavebox{\fminibox}
\newlength{\fminilength}

\def\babs{\begin{abstract}}             \def\eabs{\end{abstract}}
\def\barr{\begin{array}}                \def\earr{\end{array}}
\def\bcc{\begin{center}}                \def\ecc{\end{center}}

\def\bdes{\begin{description}}          \def\edes{\end{description}}
\def\bdoc{\begin{document}}             \def\edoc{\end{document}}
\def\ben{\begin{enumerate}}             \def\een{\end{enumerate}}
\def\beqn{\begin{eqnarray}}             \def\eeqn{\end{eqnarray}}
\def\beqnl#1{\beqn\label{#1}}           \def\eeqnl#1{\label{#1}\eeqn}
\def\beqnx{\begin{eqnarray*}}           \def\eeqnx{\end{eqnarray*}}
\def\bseqn{\begin{subeqnarray}}         \def\eseqn{\end{subeqnarray}}
\def\beq#1\eeq{\begin{equation}#1\end{equation}}
\def\bal#1\eal{\begin{align}#1\end{align}}
\def\balx#1\ealx{\begin{align*}#1\end{align*}}
\def\beqx{$$}                           \def\eeqx{$$}
\def\bfig{\protect\begin{figure}}       \def\efig{\protect\end{figure}}
\def\bfigx{\protect\begin{figure*}}     \def\efigx{\protect\end{figure*}}
\def\bfl{\begin{flushleft}}             \def\efl{\end{flushleft}}
\def\bfr{\begin{flushright}}            \def\efr{\end{flushright}}
\def\bit{\begin{itemize}}               \def\eit{\end{itemize}}
\def\bmi{\begin{minipage}}              \def\emi{\end{minipage}}
\def\bpic{\begin{picture}}              \def\epic{\end{picture}}
\def\bqu{\begin{quote}}                 \def\equ{\end{quote}}
\def\bqun{\begin{quotation}}            \def\equn{\end{quotation}}
\def\bsl{\begin{slide}}                 \def\esl{\end{slide}}
\def\btabb{\begin{tabbing}}             \def\etabb{\end{tabbing}}
\def\btabl{\begin{table}}               \def\etabl{\end{table}}
\def\btablx{\begin{table*}}             \def\etablx{\end{table*}}
\def\btab{\begin{tabular}} 
\def\btabu{\begin{tabular}}             \def\etabu{\end{tabular}}
\def\btabx{\begin{tabular*}}            \def\etabx{\end{tabular*}}
\def\bbib{\begin{thebibliography}{999}} \def\ebib{\end{thebibliography}}
\def\bver{\begin{verbatim}}             \def\ever{\end{verbatim}}

\graphicspath{{FIG/}{./}}

\def\fred #1{\textcolor{blue}{#1}}
\def\benj #1{\textcolor{brown}{#1}}
\def\phil #1{\textcolor{cyan}{#1}}
\def\rmq #1{\textcolor{red}{#1}}

\begin{document}

\tolerance=1000
\title{Double frame Tomographic PTV at high seeding densities}



\author{Philippe Cornic$^{1\dagger}$  
        Benjamin Leclaire$^{2\dagger}$
	    Fr\'ed\'eric Champagnat$^1$  
	    Guy~Le~Besnerais$^1$ 
	    Adam Cheminet$^3$  C\'edric Illoul $^2$  
	    and Gilles Losfeld $^2$
\thanks{$\dagger$ contribute equally to this paper and are co-first authors.}
}


\affiliation{
    $^1$ DTIS, ONERA, Universit\'e Paris Saclay \\
    F-91123 Palaiseau - France \\
	$^2$ DAAA, ONERA, Universit\'e Paris Saclay \\
	F-92190 Meudon - France \\
	$^3$ LMFL-Kampe, Univ. Lille, CNRS, ONERA, Arts et M\'etiers ParisTech, Centrale Lille, FRE 2017 \\
	F-59000 Lille, France \\
}


\begin{abstract}
A novel method performing 3D PTV from double frame multi-camera images is introduced. Particle velocities are estimated by following three steps. Firstly, separate particle reconstructions with a sparsity-based algorithm are performed on a fine grid. Secondly, they are expanded on a coarser grid on which 3D correlation is performed, yielding a predictor displacement field that allows to efficiently match particles at the two time instants. As these particles are still located on a voxel grid, the third, final step achieves particle position refinement to their actual subvoxel position by a global optimization process, also accounting for their intensities. As it strongly leverages on principles from tomographic reconstruction, the technique is termed Double-Frame Tomo-PTV (DF-TPTV). Synthetic tests on a complex turbulent flow show that the method achieves high particle and vector detection efficiency, up to seeding densities of around $0.08$ particles per pixel (ppp), while its root-mean-square error of velocity estimation is lower to that of state-of-the-art similar methods. Results from an experimental campaign on a transitional round air jet at Reynolds number $4600$ are also presented. During the tests, seeding density varies from $0.06$ to $0.03$ ppp on average. Associated to an outlier rejection scheme based on temporal statistics, DF-TPTV vector fields truthfully correspond to the instantaneous jet dynamics. Quantitative performance assessment is provided by introducing statistics performed by bin averaging, upon assuming statistical axisymmetry of the jet. Mean and fluctuating axial velocity components in the jet near-field are compared with reference results obtained from planar PIV at higher seeding density, with an interrogation window of size comparable to that of the bins. Results are found to be in excellent agreement with one another, confirming the high performance of DF-TPTV to yield reliable volumetric vector fields at seeding densities usually considered for tomographic PIV processing, or even higher.

\end{abstract}

\maketitle

\section{Introduction}

Tomographic Particle Image Velocimetry (TomoPIV) has been introduced as the first technique enabling the measurement of instantaneous three-dimensional (3D) velocity fields~\citep{Elsinga06}. While providing a wealth of information on a regular vector grid, allowing convenient physical analyses, one of its major limitations has been quickly identified as its more important degree of spatial filtering than in planar PIV. Indeed, due to 3D imaging constraints, the maximum acceptable seeding density maintaining an acceptable accuracy is known to be lower than for planar PIV~\citep{Scarano12,Kahler16}, imposing to perform cross-correlation with large Interrogation Volumes (IV), thereby smoothing the smallest turbulent scales.

Three-dimensional Particle Tracking Velocimetry (3DPTV) methods, on the other hand, have long been characterized by a different trade-off, as the accuracy of the velocity estimation is rather linked to the particle image size (much smaller than the typical IV size in TomoPIV), but at the cost of a low seeding density, typically lower than $0.001$ particles per pixel (ppp) \citep{Maas93,Malik93}. However, recent years have seen major improvements in 3DPTV's performance by acquiring multiple images in a time-resolved (TR) mode.

Multiframe 3DPTV, exploiting temporal consistency over a large horizon (typically 10 time steps or more) has led to obtaining reliable particle trajectories and accurate particle location, velocity and material acceleration \citep{Malik93,Schroder09,Schroder11,Schanz16,Jux18}. Associated with TomoPIV for 3D detection \citep{Schroder09,Schroder11} or with iterative stereo matching techniques like the Iterative Particle Reconstruction \cite[IPR,][]{Wieneke13} or Shake-The-Box \cite[STB,][]{Schanz16}, temporal consistency has been the key factor behind recent 3DPTV successes at seeding densities up to ppp $\approx 0.05$. However, these approaches require TR measurements, which suffer a lower signal-to-noise ratio due to limited energy per pulse, and therefore lose in accuracy in situations where good seeding or contrast quality are difficult to achieve. Furthermore, acquisition rate in a regular pulse TR mode is limited to several $kHz$, higher frequencies requiring to decreasing even more the light intensity. These techniques therefore still suffer from severe limitations in the context of high-speed flows, except if more complex and costly setups can be assembled \citep[such as a pair of interlaced double-pulse lasers enabling four pulse acquisition, see e.g.][]{Novara16}. 

In contrast, we consider here dual frame 3DPTV using a novel processing pipeline designed to achieve high performance for ppp up to 0.08. As the approach uses several algorithmic steps of the TomoPIV processing, we termed it Dual Frame Tomo-PTV (DF-TPTV). The paper is organized as follows: firstly, section \ref{sec:relwk} reviews the main recent proposed algorithms that also take as their objective to perform 3DPTV on conventional double frame acquisition to measure velocity information, and underlines the main factors defining the performance in this context. Section \ref{sec:dfptv} describes the proposed DF-TPTV method, organized in three main steps: sparse tomographic 3D particle reconstruction, temporal matching of particles, and vector refinement. Section \ref{sec:synth} then characterizes the behavior of DF-TPTV on a large range of seeding density values using synthetic data generated from a turbulent channel flow direct numerical simulation (DNS) \citep{JHU08,JHU16}. Section~\ref{sec:airjet} presents an experimental application to a cylindrical air jet. The ability of DF-TPTV to yield reliable vector fields is scrutinized both on instantaneous results, and by estimating statistical quantities which are compared with results from a classical planar PIV system. Section~\ref{sec:conclusion} is devoted to conclusions and perspectives.

\section{Related works}\label{sec:relwk}

Recently, high accuracy measurements have been obtained from 3DPTV by \citet{Fuchs16} and \citet{Aguera16}, with performance illustration by computing ensemble statistics through spatio-temporal binning of 3DPTV vectors. However, these methods differ significantly in their processing steps.

\citet{Aguera16} first conduct 3D particle detection by a classical 2D particle detection in the images and stereoscopic triangulation, and then solve the temporal matching in two steps. Particles at the first time instant are displaced using a "predictor" motion field obtained by correlation on low-resolution TomoPIV volumes. They are then matched with particles at the second time instant by nearest-neighbor association. The method appears to be limited to low seeding densities (ppp $=0.003-0.005$) setting hard constraints either on the size of the bin, or on the number of acquired snapshots, in the context of statistical estimation by bin averaging. The authors use as a matter of fact large bins and propose a technique to mitigate the influence of a mean velocity gradient inside each bin that otherwise would bias the computation of second order statistics.

\citet{Fuchs16} reconstruct a volume by TomoPIV (using either MLOS or MART) and detect 3D particles by fitting 3D Gaussian to voxel intensities. A selection of particles is then applied: a detection is confirmed if its projections in PIV images can be associated unambiguously to a unique particle image in each frame. This rule eliminates almost all ghosts at the cost of loosing a significant number of true particles. Selected detections are then triangulated and propagated to the next time instant by means of a previously estimated displacement field. They finally use a matching process which takes into account the spatial regularity of the motion field so as to reduce outlier vectors as much as possible. Their method has been applied to the computation of mean flow velocity and Reynolds stresses of a turbulent boundary layer, with results equivalent to 3DPTV on TR data \citep{Fuchs16}.

For exhaustiveness, let us finally quote that a dual frame version of Shake-The-Box \citep[STB,][]{Schanz16} has been successfully applied to case C of the 4th PIV Challenge in 2014~\citep{Kahler16}. STB relies on a specific combination of prediction/correction/matching steps. In particular, accurate prediction is enabled by long particle tracks in time resolved mode. In the context of challenge C of the 4th PIV Challenge, a null velocity predictor was used and yielded good results, because the true displacement was small enough.

Both the mentioned methods, \citet{Aguera16} and \citet{Fuchs16}, are characterized by a relatively low maximum density of estimated 3D vectors, as a consequence of their choices of 3D reconstruction method. The epipolar stereo matching method used in \citet{Aguera16} can only work at very low seeding densities (ppp $=0.003-0.005$); TomoPIV methods used in \citet{Fuchs16} yield a high percentage of ghost particles leading the authors to choose a drastic selection rule. 

In the following, we show that, in contrast, by using in particular sparse TomoPIV reconstruction, \citep{Cornic15} we have been able to contain the proportion of ghosts in the 3DPTV process while maintaining a higher number of estimated vectors. 

\section{The Double Frame Tomo-PTV technique}\label{sec:dfptv}

\begin{figure*}
\begin{center}\leavevmode
\includegraphics[width=1.5\columnwidth]{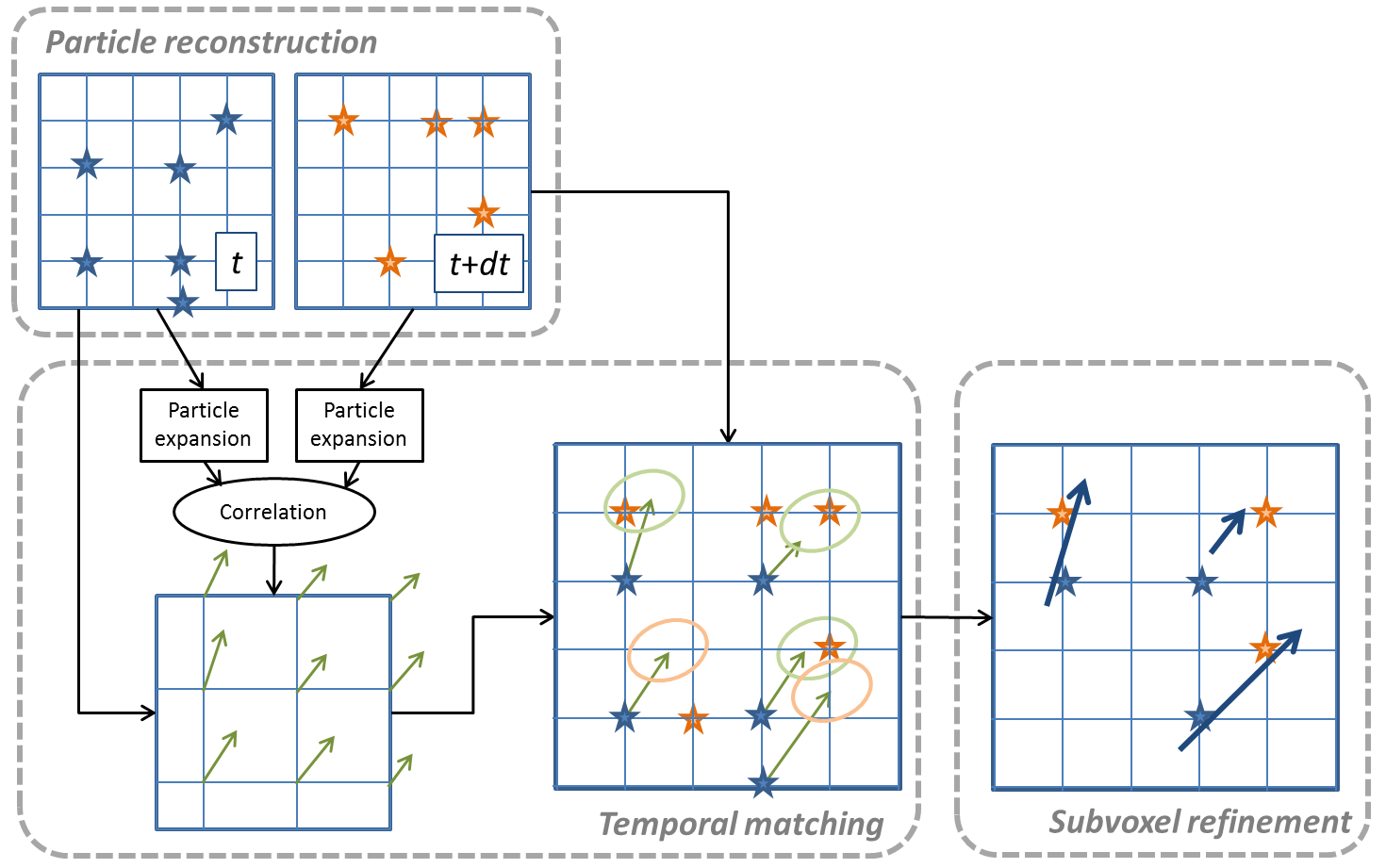}
\caption{Method overview (illustrated in 2D): Particle reconstructions at $t$ and $t+dt$ (yielding sparse/spiky particle representations in the voxel space), temporal matching, and subvoxel refinement of the matched particles. The temporal matching comprises three steps: estimation of an initial displacement field by 3D correlation on a coarse grid of the initially expanded particles, application of the interpolated displacement field to particles at time $t$ to predict their position at time $t+dt$, and local matching with particles at time $t+dt$. }
\label{fig:3DAdvection}
\end{center}
\end{figure*}

The technique involves 3 stages, sketched in Figure~\ref{fig:3DAdvection}: initial particle reconstructions on a voxel grid at the two instants, 3D matching of particles yielding a first estimate of the displacement vectors, and subvoxel refinement of the particle positions and thus of the displacements. 

Note that the very first version of the technique was introduced in \citet{Cornic14}, with good performances at densities up to ppp $\approx 0.03$, as attested by the 4th PIV challenge \citep{Kahler16}. It was further improved in \citet{Cornic15b}. The following presentation 
corresponds to the stabilized and optimized version of the method, featuring in particular much simplification compared to our previously communicated work. 

\subsection{Initial particle reconstructions}
The first step is a fast and efficient localization and intensity reconstruction of particles on a 3D voxel grid based on sparsity principles, comprehensively described in \cite{Cornic15}.  
It is applied to obtain separate and initial particle reconstructions at the two time instants, in the form of indices and intensities of voxels supposed to contain a particle.
The initial step is a traditional MLOS operation \citep{Atkinson09} on a grid with a voxel-to-pixel ($v/p$) ratio of 0.5. The number of potential particles is further reduced by retaining only voxels corresponding to local intensity maxima ("LocM" strategy). The tomographic reconstruction then relies on the Particle Volume Reconstruction (PVR) imaging model, which seeks to explain the images of a particle with a cluster of few non zero voxels \citep{Champagnat14}. The PVR system $Y=WE$ relates pixel intensities $Y$ to voxel intensities $E$ through a weight matrix $W$ made of Point Spread Function (PSF) samples. A sparse solution is defined through the following constrained minimization problem:
\begin{equation}
\underset{E}{\min}\left \|Y-WE  \right \| \mbox{subject to} \left \| E  \right \|_0 \leq S , 
\label{eq:cosamp}
\end{equation}
where $\left \|E  \right \|_0$ is the number of non zero entries of $E$. We use CoSaMP~\citep{Needell08}, a sparsity based algorithm to solve this problem over voxels yielded by the LocM selection. The main parameter is $S$, the upper bound on the number of non zero voxels in the reconstruction, which in practice is taken as the expected maximum number of particles in the volume.
The overall reconstruction algorithm is termed LocM-CoSaMP. As shown by \citet{Cornic15}, it has a high efficiency to preserve real particles and remove ghosts, which, as discussed earlier, is a critical asset in the context of dual frame 3DPTV.

\subsection{3D particle temporal matching}
This step consists in identifying the same physical particles in the two instants, with the aim of reducing as much as possible the number of ghosts in the individual reconstructions. As proposed in previous works, we proceed in two steps, first using a predictor motion and then a nearest-neighbor association restricted to a small region around the predicted position. In a preliminary version of the algorithm~\citep{Cornic14}, 2D displacement fields in the four images were used as motion predictor, but this was found to be insufficient, in particular for turbulent flows. A 3D motion predictor derived from the correlation of the two reconstructed volume was introduced in \citet{Cornic15b} and also used in \citet{Aguera16}. 

In practice, LocM-CoSaMP reconstructions are first post-processed before correlation. Non zero voxels are expanded with Gaussian filtering on a low-resolution, voxel-to-pixel ratio $v/p \geq 2$ grid, on which the 3D correlation will be performed. Gaussian filtering is necessary in the grid transfer process, as the PVR model, used in LocM-CoSaMP, is designed to yield spiky particle reconstructions, extending to a minimum number of voxels \citep[instead of the more traditional "blob" paradigm of 3D PIV, see][]{Champagnat14}. 3D correlation is obtained using FOLKI3D \citep{Cheminet14}, a 3D extension of FOLKI-PIV \citep{Champagnat11}. Once particles are propagated, a search region of matches of three-voxel radius (expressed here in $v/p=1$ units) is in practice sufficient.

\subsection{Subvoxel refinement}

After matching, the obtained particles are still located on a voxel grid, so that a final step performing subvoxel localization is required. Contrary to techniques such as the Iterative Particle Reconstruction \citep[IPR,][]{Wieneke13} or STB \citep{Schanz16}, that process all the particles sequentially, subvoxel refinement is here performed through a global optimization so as to fully account for the interactions between the particles in the images. 
The objective function to minimize is the sum of squared differences (SSD) between the recorded images and the images corresponding to the projections of the obtained 3D particles:
\begin{equation}
    J({\Xb_p,E_p}) =   \sum_j \sum_x \left\|Y_j(x) - \sum_p E_p h(x-F_j(\Xb_p)) \right\|^2,
    \label{eq:bundle}
\end{equation}
where $Y_j$ are the recorded images, $x$ a given pixel coordinate in an image, $F_j$ is the projection function in image $Y_j$ yielded by the calibration and $h$ is the PSF. This SSD is thus performed over all pixels of all images.
Without loss of generality and to alleviate the notations, we suppose that $h$ is constant and the same for all images. 

The non linear least squares criterion $J$ is minimized over the 3D positions and intensities $\{\Xb_p,E_p\}$ of the particles, i.e. potentially a huge number of variables. To cope with this issue, we used the L-BFGS algorithm \citep{Nocedal}.
We perform this optimization independently at time steps $t$ and $t+dt$, on matched particles only. In other words, we move independently the two ends of each 3D vector, as illustrated in Fig.~\ref{fig:3DAdvection}. At the end of the process, DF-TPTV thus produces estimated 3D displacement vectors with real, subvoxel coordinates.

\section{Assessment on synthetic data}\label{sec:synth}
Being a PTV algorithm, the method must be assessed for performance both by the number of vectors produced, compared to the number of tracers present in the observation volume, and by their precision, in terms of RMS error on the 3D positions and displacements. This is the purpose of the simulation study presented in this section. We here intend to characterize the behavior of DF-TPTV with respect to seeding density, over a large range of ppp. In doing so, we also provide some elements of comparison with performances reported in the literature.

\subsection{Synthetic data generation}

\begin{figure}[htbp]
\begin{center}\leavevmode
\includegraphics[width=\columnwidth]{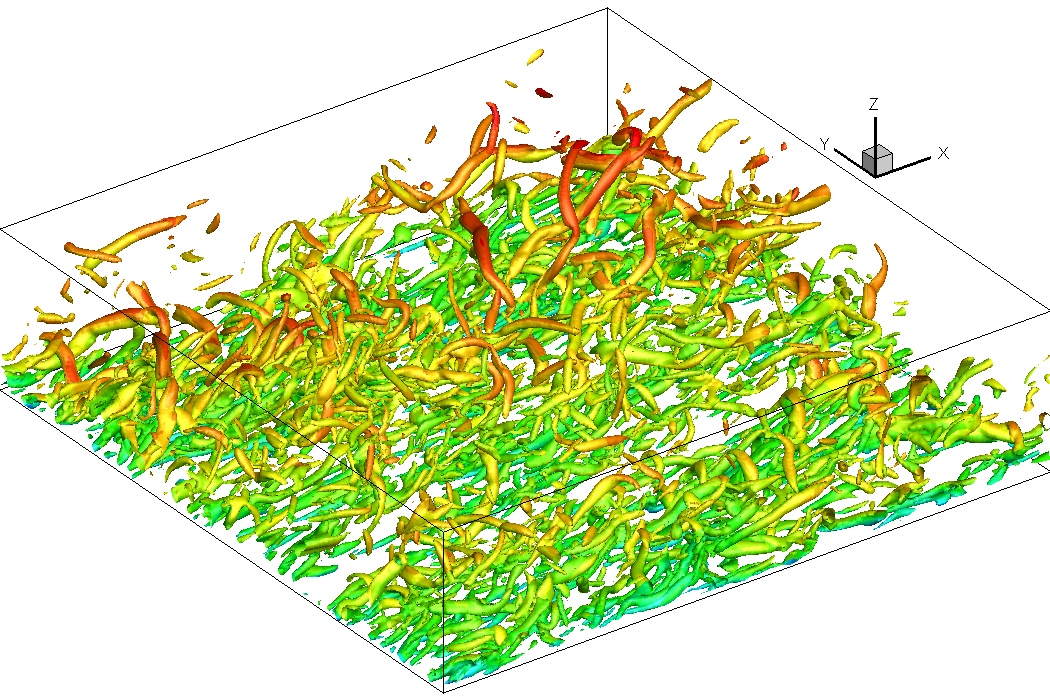}
		\caption{Flow snapshot used for synthetic data generation  (iso-contours of Q-criterion color-coded by velocity norm), obtained from the Johns Hopkins Turbulence Database \cite[channel flow case, ][]{JHU08,JHU16}. Friction Reynolds number is equal to $Re_{\tau}=1000$. The observed volume has a thickness of $400$ wall units starting from the wall, and a length and width of $2000$ and $2200$ wall units, respectively.
	\label{fig:jhuflow}}
\end{center}
\end{figure}

Similar to e.g. case D of the 4th PIV Challenge \citep{Kahler16}, we use here one of the flow cases of the Johns Hopkins Turbulence Database, namely the turbulent channel flow DNS \citep{JHU08,JHU16}. We define $x, y$ and $z$ as the streamwise, spanwise and wall-normal directions, respectively. A domain $100\times 110 \times 20$ mm$^3$  has been simulated, taken in contact with the lower wall. Figure \ref{fig:jhuflow} shows iso-contours of the Q-criterion color-coded by the local velocity norm of the velocity snapshot used for particle displacement, illustrating its complex turbulent structure. As the database uses dimensionless coordinates, a scaling has been chosen, such that the size of one voxel roughly corresponds to the viscous length scale of the database, i.e. that the first voxel away from the wall corresponds to one wall unit. This leads to a voxel size of $50~\mu{m}$, and a volume extension of $2000\times{2200}\times{400}$ wall units.

A traditional, four cameras observation setup is simulated. With the world coordinate origin located on the wall of the channel, at the middle of the illuminated zone in the streamwise and spanwise directions, these cameras are located at the corners of the base of a right square pyramid, whose apex is at $(0,0,0)$ and height coincides with the
$z$ axis. All four cameras have a  $2016\times 2016$ pixel sensor, with a pixel pitch of $11\mu$m (e.g. similar to the PCO Dimax S4), and are equipped with a lens of focal length $f=200$ mm. Their roll position with respect to the optical axis, and their Scheimpflug angle, are computed by assuming that they are all in focus at mid-thickness of the illuminated volume. Imaging of the particles is supposed to be diffraction limited. The apparent diameter of a particle's image is $d_\tau = 2.4$ pixels, resulting from a Gaussian PSF of standard deviation $\sigma=0.6$ pixel integrated over the pixel surface. No noise is added on the camera sensor.

Laser illumination is modeled as a constant intensity throughout the illuminated volume, whose extension is infinite in the $y$ direction, and spans the range $[-50\, 50]$ mm and $[0\, 20]$ mm in the $x$ and $z$ directions, respectively. For all synthetic experiments, the upper bound $S$ on the LocM-CoSaMP sparse reconstruction (see Eq.~\ref{eq:cosamp}) is chosen equal to the true number of particles considered in the simulation and  
seen simultaneously by the four cameras. It ranges from 33,280 (ppp $=0.01$) to 332,287 (ppp $=0.1$). As described in more detail in \cite{Cornic15}, in a real experiment --- as the one considered in Section \ref{sec:airjet} --- parameter $S$ is deduced in practice from an estimation of the image seeding density.

\begin{figure*}
\begin{center}\leavevmode
\includegraphics[width=.85\columnwidth]{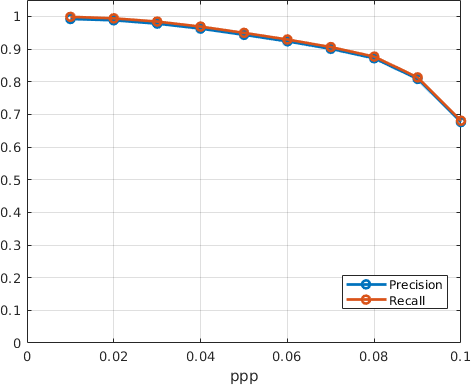}  \includegraphics[width=.85\columnwidth]{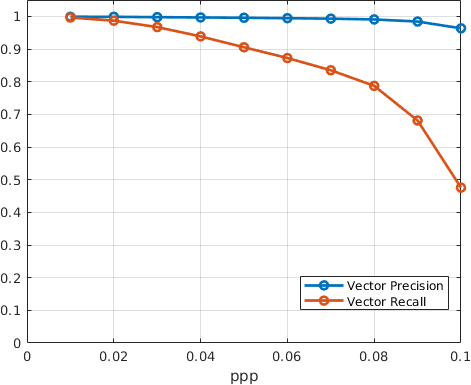}  		\caption{Detection performances of DF-TPTV with respect to seeding density (ppp). Left: Precision and Recall of the particles reconstruction step. Right: Precision and Recall on velocity vectors. }
	\label{fig:precrec}
\end{center}
\end{figure*}

\subsection{Detection performances}\label{sec:perfodetection}
The set of particles seen by all the cameras is used as the ground truth for measuring the performance of DF-TPTV in terms of particle detection. To quantify it, we adopt the metrics of 
\citet{Champagnat14} and \citet{Cornic15}. A detection is referred to as a true positive (TP) if it lies within $1$ voxel chessboard distance 
of a true particle. Otherwise it is a false positive (FP), a.k.a a ghost. A non-detection, or false negative (FN), is reported for a true particle with no detection within $1$ voxel distance. From these basic quantities two indices of performance are defined:
\begin{equation}
    \text{Recall} = \frac{\#\text{TP}}{\#\text{TP}+\#\text{FN}} \,\mbox{and}\, \text{Precision} = \frac{\#\text{TP}}{\#\text{TP}+\#\text{FP}}, 
\end{equation}
where \# stands for number of. Recall is the number of detected particles over the total number of true particles in the volume, and is thus the detection rate. Precision is the fraction of true particles among the detected particles. The best achievable performance is given by Recall $=1$ (\#FN $=0$, every particle is detected) and Precision $=1$ (\#FP $=0$, no ghosts).

Figure \ref{fig:precrec} left gives the Precision and Recall of the reconstruction at initial time $t$, as a function of the seeding density expressed in particles per pixel. As LocM-CoSamp is tuned to retrieve the exact number of particles, one has $S$=\#\text{TP}+\#\text{FN}=\#\text{TP}+\#\text{FP}, so that Recall and Precision are equal. Performance is observed to decrease first slowly with the seeding density; value ppp $=0.08$ then sets a break after which the performances decline faster. Note that such a concentration is already above the usually acknowledged optimum for TomoPIV, i.e. roughly around $0.05$ ppp. As mentioned in section \ref{sec:relwk}, the selection rules adopted by \citet{Fuchs16} led them to a quasi negligible percentage of ghosts, at the cost of low Recall (detection rate). According to \cite{Fuchs16}'s Fig. 1(a), their Recall is around 70\% at ppp $=0.05$, while we here obtain a  95\% Recall at this ppp. Even though the dataset considered by \cite{Fuchs16} and the present one are different and therefore cannot be compared directly, these figures tend to support the fact that the two methods rely on a different operating point between Precision and Recall. 
In other words, the reconstruction step of DF-TPTV appears more balanced between the percentage of ghosts and the detection rate compared to \citet{Fuchs16}. We will see below that the fact of tolerating a higher rate of ghosts allows here to retrieve a higher percentage of true vectors while maintaining the ratio of outliers vectors as low as a few percent.

Precision and Recall can also be computed on velocity vectors. Here, the ground truth is made of all vectors defined by a true particle visible at the two time instants by all cameras. A true positive vector (TPv) stems from the detection of the same particle at both instants and the correct matching of the two detections.
The fraction of true positive vectors TPv among all retrieved vectors is the Precision, while the fraction of TPv among all true vectors is the Recall. Figure \ref{fig:precrec} (right) shows vector Precision and Recall of DF-TPTV over a range of ppp up to $0.1$. Upon comparing for each ppp the values of Recall on figures \ref{fig:precrec} left and right, one observes that the obtained vector Recall is approximately equal to the square of the reconstruction Recall. This means that vector Recall is mainly limited by the missing detections rather than by the matching step, and that the latter appears close to the optimum, justifying the use of a predictor motion field based on 3D correlation. Moreover, contrary to the Precision of particle detection, vector Precision stays close to $1$ when the ppp increases. This means that the 3D matching process filters out most of the ghost particles.

The vector Recall can be translated straightforwardly to 
the \textit{effective amount of particles by pixel} metric introduced by \citet{Fuchs16}, by multiplying it with the seeding density. For instance, at ppp equal to $0.05$, Fig. \ref{fig:precrec} right indicates a vector Recall equal to $0.9$, thus corresponding to a $0.045=0.05 \times 0.9$ effective amount of particles by pixel. As a comparison, \cite{Fuchs16} report a maximum effective amount of $0.032$ (see their Fig. 2b) --- however bearing in mind the slightly different characteristics of their dataset. The percentage of ghost vectors among the total number of true vectors can be computed as
\begin{equation}
    \mathrm{\% ghost}= \frac{(1-\mathrm{Precision})\times \mathrm{Recall} }{\mathrm{Precision}}.
\end{equation}
For the considered ppp $=0.05$, Figure \ref{fig:precrec} right shows that $\mathrm{Recall}=0.9$, $\mathrm{Precision}=1-0.01$, consequently the percentage of ghost is equal to $0.9\%$, i.e. indeed a very low proportion of the vector field.

\subsection{Position and velocity accuracy}\label{sec:accuracy}

\begin{figure*}
\begin{center}\leavevmode
\includegraphics[width=.85\columnwidth]{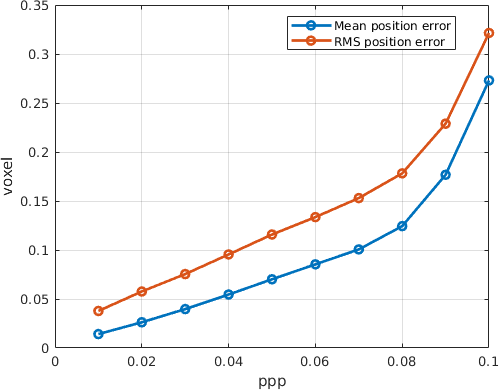} 
\includegraphics[width=.85\columnwidth]{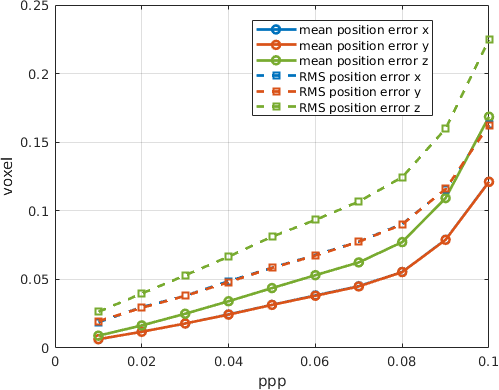}
		\caption{Precision of particle localization of the DF-TPTV method, expressed in voxel units, as a function of the seeding density. Left: total mean and RMS position error. Right: mean in absolute value and RMS position error for each displacement component.}
	\label{fig:locerror}
\end{center}
\end{figure*}

\begin{figure*}
\begin{center}\leavevmode
\includegraphics[width=.85\columnwidth]{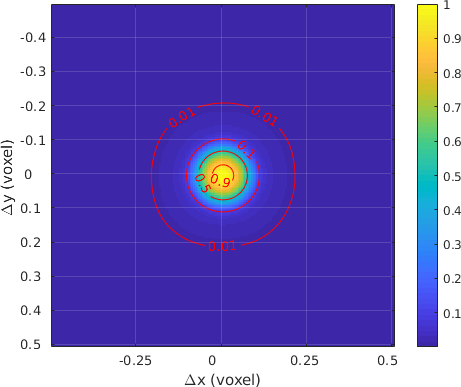} 
\includegraphics[width=.85\columnwidth]{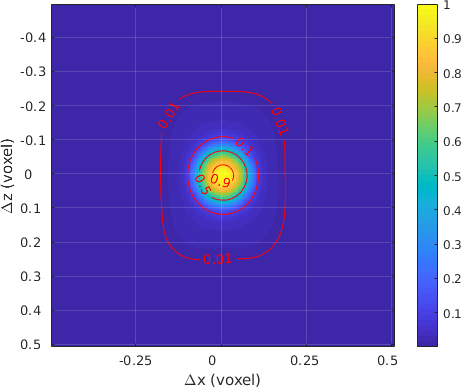}
		\caption{Position error probability density function for DF-TPTV at ppp=0.05. Left: $xy$-plane. Right: $xz$-plane.}
	\label{fig:pdf}
\end{center}
\end{figure*}

Figure \ref{fig:locerror} left shows the mean (blue) and RMS (red) location error in voxel units associated to the correctly detected particles (TP). Figure \ref{fig:locerror} right shows the repartition of these errors in absolute value on the $3$ displacement components (note that $x$ and $y$ curves collapse). Both are expressed in voxel units. These curves show a gradual decay of precision with increasing ppp, until ppp $=0.08$ where a change in slope occurs. Not surprisingly, error on the depth component $z$ is higher. It is worthwhile noting that the mean location errors are lower than that reported in the literature for state-of-the-art particle reconstruction or double frame 3D PTV techniques, \eg IPR  \citet{Wieneke13} or \citet{Fuchs16}, and comparable to the 3D positional errors of STB obtained for the first instants of a given sequence \citep{Schanz16}. Figure \ref{fig:pdf}  displays the location error probability density function (pdf) in the $xy$ and $xz$ planes, for ppp $=0.05$. It may be seen that the pdf decays very fast in the vicinity of the origin and is elongated in the $z$ direction in the $xz$ plane, in line with the higher error values corresponding to this component.

\begin{figure}[htbp]
\begin{center}\leavevmode
\includegraphics[width=\columnwidth]{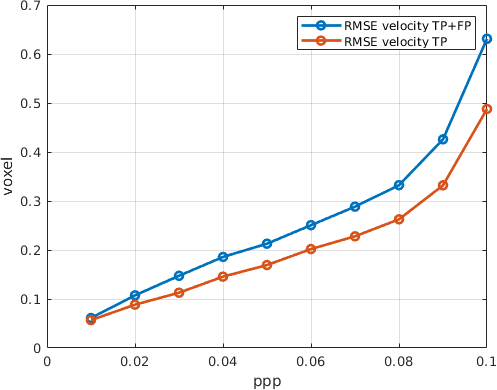}
		\caption{RMS error on velocity, $RMS_v$, computed over all vectors (blue) and only true positive vectors (red) as a function of seeding density.}
	\label{fig:velerror}
\end{center}
\end{figure}

Vector fields obtained by DF-TPTV are assessed using the vector RMS (RMS$_v$), measuring the error between estimated and ground truth displacements of the detected particles, and defined as:
\begin{equation}
    \mbox{RMS}_{v} = \sqrt{\frac{1}{P}\sum_{p=1}^{P} \sum_{i=1}^{3} (u_i(x_p,y_p,z_p)-\Tilde{u}_i(x_p,y_p,z_p))^2}, 
\end{equation}
where $P$ is the number of detected particles, $(u_1,u_2,u_3)$ is the ground truth 3D displacement and $(\Tilde{u}_1,\Tilde{u}_2,\Tilde{u}_3)$ is the DF-TPTV estimation of the 3D displacement. Both velocities are evaluated at locations $(x_p,y_p,z_p)$ of the detected particles (and not at the locations of the ground truth particles). Figure \ref{fig:velerror} shows RMS$_{v}$ as a function of the ppp computed over all the vectors (blue curve) and only over the TP vectors (red curve). Trends are logically similar to that observed for particle localization error, with a gradual increase of RMS$_{v}$ up to ppp $=0.08$, and a more pronounced performance loss for higher densities. It can also be seen that, although there are only few wrong vectors compared to the number of good ones, they have a noticeable influence on the RMS when the ppp increases. This will motivate the introduction of outlier rejection in the processing of experimental data, in section \ref{sec:airjet}.

\section{Experimental results on a round air jet}\label{sec:airjet}
\subsection{Experimental setup}
\label{sec:expesetup}

\begin{figure*}[htbp]
\begin{center}\leavevmode
\includegraphics[width=0.8\textwidth]{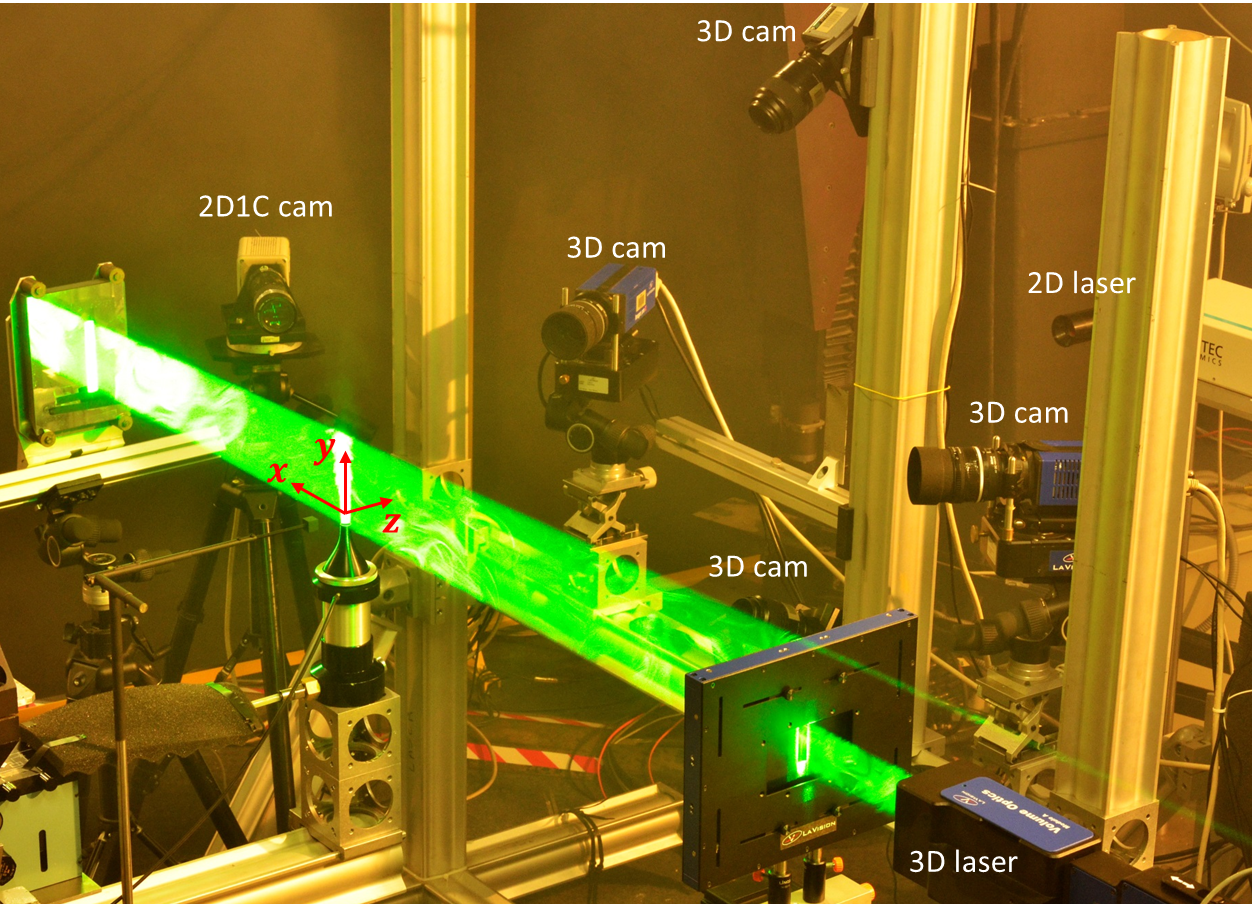}
		\caption{Transitional jet experiment, featuring 3D PTV measurement in a parallelepiped of $16~mm$ thickness in the $z$ direction (with volumic illumination on), observed by four cameras, and planar PIV measurement in a jet-longitudinal $yz$ plane (camera in perspective observation, therefore used as a reference for the streamwise velocity component only).}
	\label{fig:expepiv}
\end{center}
\end{figure*}

Demonstration on experimental data is performed by considering a round air jet of exit diameter $D=12~mm$. With $V_j=5.8~m.s^{-1}$ the velocity at the centre of the exit plane, the flow Reynolds number based on $D$ and $V_j$ is equal to $Re=4600$. Flow conditions in the exit plane are transitional/turbulent in the boundary layer, due to the presence of a small circular step at the nozzle wall $20~mm$ upstream from the exit. In the following, only dimensionless quantities, built using $D$ and $V_j$ as reference length and velocity, are considered. The centre of the jet exit plane is taken as the origin O of the coordinate system; $y$ denotes the direction aligned with the jet axis, here vertical, and $x$ and $z$ the horizontal axes (see figure \ref{fig:expepiv}).

The near field of this jet, up to $y/D\approx 7.3$, is measured using two PIV systems. 3D measurements are acquired in a parallelepiped with its largest edges in the $x$ and $y$ directions, with an approximate thickness of $16~mm$, centered around O in the $z$ direction (see figure \ref{fig:expepiv}). Illumination is achieved using a Quantel Twin Ultra Nd-Yag laser delivering $120$mJ per pulse, and observation by two Dantec HiSense and two LaVision Imager ProX 4 Mpixels cameras ($2048\times 2048$ pixel), set up in a cross like configuration. A mirror is placed in order to reflect back the volumic illumination and compensate for the different scattering condition among the cameras. 

In order to have a reference on flow quantities for performance assessment, and also to provide an independent seeding density evaluation during the tests, an additional planar PIV measurement system is set up. It is made of one Litron Laser Dual Nd-YAG $532$nm laser and one Dantec HiSense 11M ($4000 \times 2672$ pixel) camera. The laser sheet, of estimated thickness $1.5~mm$, is located in the $x=0$ plane, orthogonal to the main axis of the tomographic laser. Due to the presence of the volumic illumination, the camera optical axis cannot be placed perpendicular to the laser sheet, and is thus set with a slight perspective. The camera is equipped with a Scheimpflug mount, and calibration is applied in order to compensate for perspective distortion in the observation. Due to this single-camera observation, this system is used for flow comparisons only on the streamwise, $v$ component, and its camera is therefore labelled as "2D1C camera" in figure \ref{fig:expepiv}. 

Acquisitions of the 3D and planar systems are intertwined, in the sense that during a run, 3D and planar snapshots are acquired alternatively, with a separation time of $0.25$s. This results in a respective acquisition frequency for both systems of $2$Hz. Besides, both also operate with the same inter-frame time of $50\mu$s.       

\begin{figure*}[htbp]
\begin{center}\leavevmode
\includegraphics[width=0.9\textwidth]{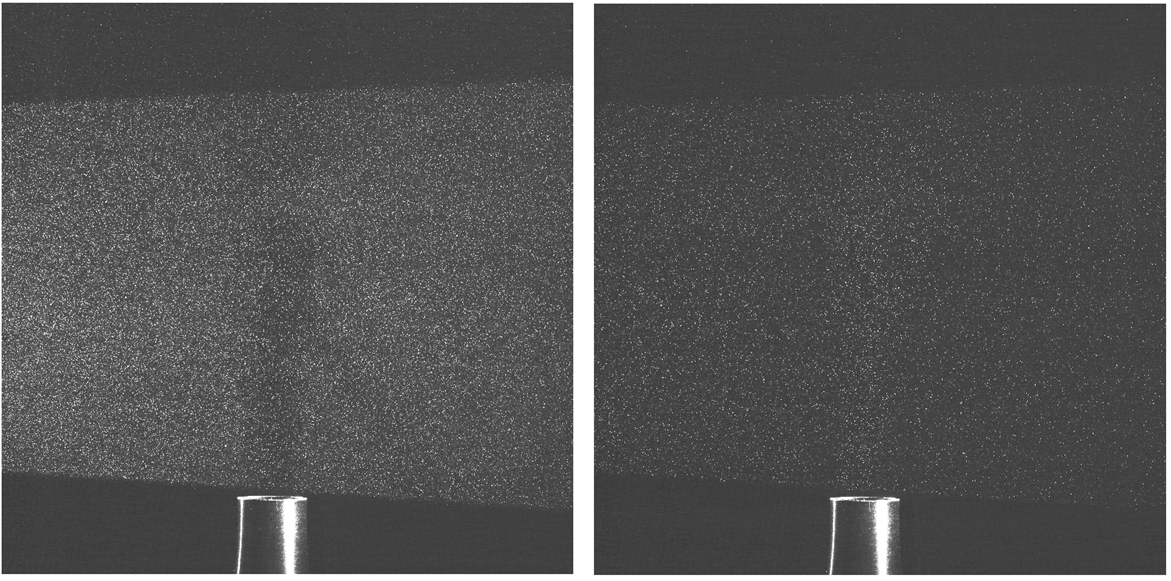} 
		\caption{Sample particle images from one of the 3D cameras, average seeding density estimated to $0.06$ ppp (left) and $0.03$ ppp (right). Contrast has been enhanced for display purposes.}
	\label{fig:Sample3Dimages}
\end{center}
\end{figure*}

Seeding is achieved using two different aerosol generators producing DEHS droplets, one whose particles are injected into the jet settling chamber (thus seeding the jet), and one used to seed the experimental room (and thus the outer shear layers and the entrained flow). Whereas the former continuously injects particles during the run, the latter is operated by initially saturating the (quite large) room, and waiting for homogenization to begin acquisition, without further injection later on. Consequently, as illustrated in figure~\ref{fig:Sample3Dimages}, external seeding progressively decays during the run. Overall, when considering the global duration of the run presented here (of the order of $8$ minutes, corresponding to $1044$ snapshots), the jet is observed to be on average more densely seeeded than the ambient air, and the global seeding density to progressively decrease from $0.06$ to $0.03$ ppp. 

\subsection{Processing parameters}
\label{sec:acqprocpar}
Calibration and self-calibration of the 3D system are done according to a pinhole model \citep{Cornic16}, leading to a voxel size ($v/p=1$) of $53.8\mu$m. Standard pre-processing steps are applied to the images of the 3D cameras before applying the DF-TPTV method, i.e. historical minimum subtraction and thresholding by identifying the average noise level from a non-illuminated zone (see figure \ref{fig:Sample3Dimages}). Additionally, to compensate for differences in dynamics between the camera images (due to the use of different camera models, and different scattering conditions in spite of the mirror), normalization was also necessary to obtain comparable signal-to-noise ratios between the cameras.

To obtain velocity fields from these pre-processed data, the DF-TPTV method is set by choosing the number of particles to retrieve (sparsity parameter $S$ in the LocM-CoSaMP particle reconstruction step) using the ppp estimation from the 3D images, which is found to be close to the estimation yielded by the planar PIV system. Other tuning parameters, i.e. pertaining to matching and subvoxel refinement, are left to their default values, mentioned in section \ref{sec:synth}.

In order to filter the results from remaining outliers, we introduce a rejection post-processing. To do so in a most adapted and efficient way, following e.g. \cite{Griffin10}, we choose to rely on temporal statistics computed by bin averaging, that will be introduced to compute mean flow fields (see section \ref{sec:StatRes}). As already noted in the PTV literature, such an approach, when available, is more efficient for turbulent flows than relying on comparisons to a spatial neighborhood. Contrary to \cite{Griffin10} however, we here rely on a simpler combination of univariate statistical rejection rules, as we choose to reject a given vector if any of its $(u,v,w)$ components deviates from more than three standard deviations from its mean (the latter two referring to that of the bin to which the vector belongs). 

Finally, note that variations in illumination within the volume led in practice to restrict the results to $-0.61\leq z/D \leq 0.40$, i.e. a $13~mm$ thickness, in order to exclude edge effects where light intensity was lower. More details on this are given in section \ref{sec:AvgCarac}. Instantaneous fields will also be restricted to this zone for consistency. Also, note that the most upstream location of the volumic vector fields yielded by DF-TPTV is located slightly above the jet exit, i.e. $y/D\approx{0.3}$. As can be seen in figure \ref{fig:Sample3Dimages}, in order to avoid intense light reflections on the nozzle, the laser volume was indeed positioned slightly above its exit plane.

\subsection{Instantaneous results}
\label{sec:InstRes}

As a first experimental illustration of the DF-TPTV method performance, we show in figure \ref{fig:Snapshots} the instantaneous vector field obtained for the first pulse of the run, at highest seeding density, estimated to around $0.06$ ppp (see also corresponding particle images of one of the cameras in figure \ref{fig:Sample3Dimages} left). We show both raw results and results post-processed via the outlier rejection method presented above, in order to illustrate its effect. Out of the $30,787$ vectors obtained in the retained $-0.61\leq{z}\leq{0.40}$ zone, outlier rejection filtered roughly $2.7\%$ of them, leading to a useful set of $29,947$ vectors. This post-processing seems indeed to be efficient and adapted, rejecting a large number of spurious vectors while keeping the physical ones, as can be observed most evidently in the external flow.

\begin{table*}
\begin{tabular}{l}
\includegraphics[width=\textwidth]{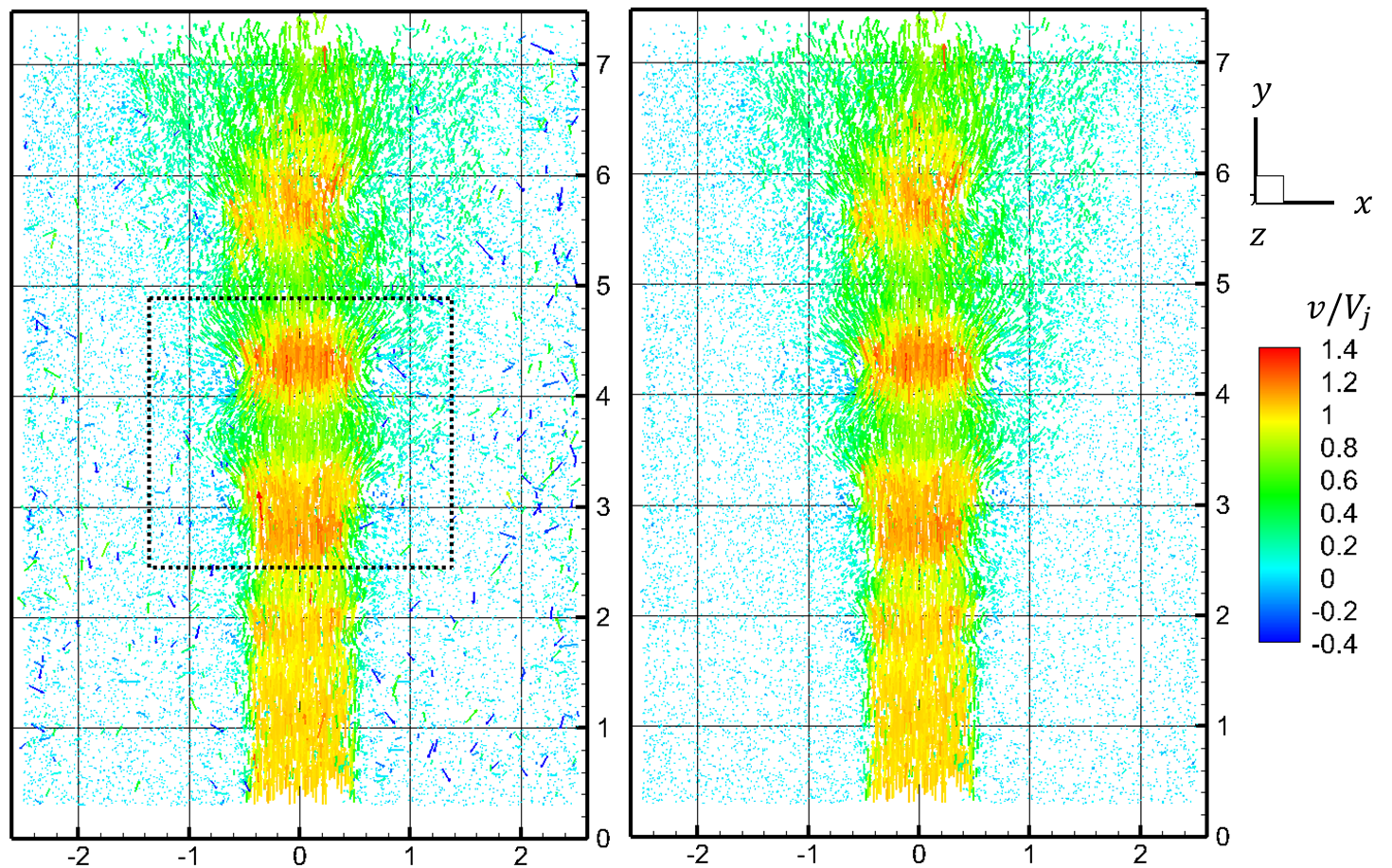}       \\
\hspace{0.2cm}\includegraphics[width=0.85\textwidth]{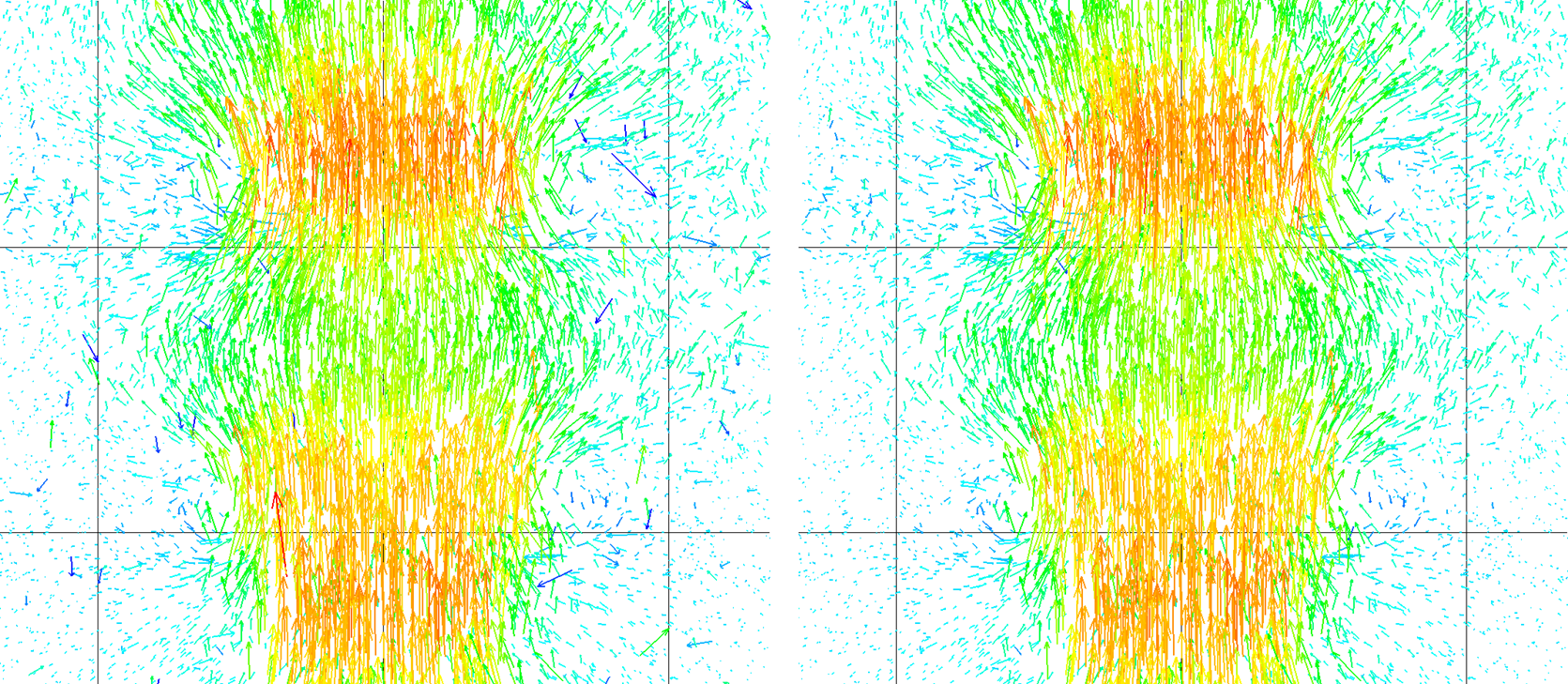}
\end{tabular}
\captionof{figure}{Instantaneous vector field of the transitional $Re=4600$ round air jet yielded by the DF-TPTV method, color-coded by the normalized streamwise component $v/V_j$. Left: raw result, right: result with outlier filtering (see text for details). Full field (top) and close-up on the region delineated by the dashed rectangle.}
\label{fig:Snapshots}
\end{table*}

Figure \ref{fig:Snapshots} also shows that the expected instantaneous structure of this transitional jet is successfully retrieved by the DF-TPTV. Starting from the jet exit, one first observes uniform axial velocity in the jet core ($|x|/D<0.4$, $y/D$ ranging from $0.3$ to around $1.5$), together with a thin shear layer on the lateral edges. Note that in this jet potential core zone, as the volume spans over $-0.61\leq{z}\leq{0.40}$, one also observes the start of the shear layer in nearly all azimuthal directions, which translates here into the presence of vectors also with lower axial velocities (of around $v/V_j\approx{0.5}$, appearing in green). Following the jet evolution in the downstream direction, one then observes typical toroidal vortical structures in the shear layer, due to the Kelvin-Helmholtz instability, coinciding with accelerations in the jet core. As also expected, in each region separating two successive of these vortices, the jet column is seen to expand, leading to flow deceleration on the axis as a result of mass conservation. These typical dynamical features can be observed in more detail in the close-up also shown in figure \ref{fig:Snapshots}.

\subsection{Statistical results}
\label{sec:StatRes}

For further performance assessment, we now consider mean and fluctuating velocities in the jet obtained by statistical averaging, which we compare to the same quantities yielded by planar PIV. We decompose each velocity component as, for instance on the axial one, $v=V+v'$, where $V$ denotes the mean and $v'$ the instantaneous fluctuation.

\subsubsection{Runs and averaging characteristics}
\label{sec:AvgCarac}
\begin{table*}
\caption{Characteristics of mean flows yielded by DF-TPTV and planar PIV:  spatial resolution (interrogation window or bin size) and number of samples available for averaging. Planar PIV results correspond to a separate run performed at higher seeding density.\label{tab:AveragingResSamples}}
\begin{center}
\begin{tabular}{|l|c|c|}
\hline
\bf{Parameter} & \bf{DF-TPTV} & \bf{Planar PIV} \\
\hline
Horizontal ($x,z$) / radial resolution ($D$) & From $0.09$ ($r=0$) & $0.014$ \\
& to $0.01$ ($r\geq{0.4}$) & \\
\hline
Vertical ($y$) resolution ($D$) & $0.084$ & $0.084$ \\
\hline
Number of snapshots & $1044$ & $3000$ \\
\hline
Number of samples for averaging & From $\approx 2200-2300$ ($r=0$) & $3000$ \\
& to $\approx 1000-1200$ ($r\geq{0.4}$) & \\
\hline
\end{tabular}
\end{center}
\end{table*}

As 3D PTV and planar PIV operate at different optimal seeding densities, a dedicated run at higher seeding density was performed in order to obtain the reference mean flow with planar PIV. We thus here present the compared characteristics of the runs for 3D-PTV and planar PIV, as well as the respective methods for obtaining statistics, and their spatial resolution. Table \ref{tab:AveragingResSamples} sums up associated relevant quantities.
\begin{figure*}[htbp]
\begin{center}\leavevmode
\includegraphics[width=0.9\textwidth]{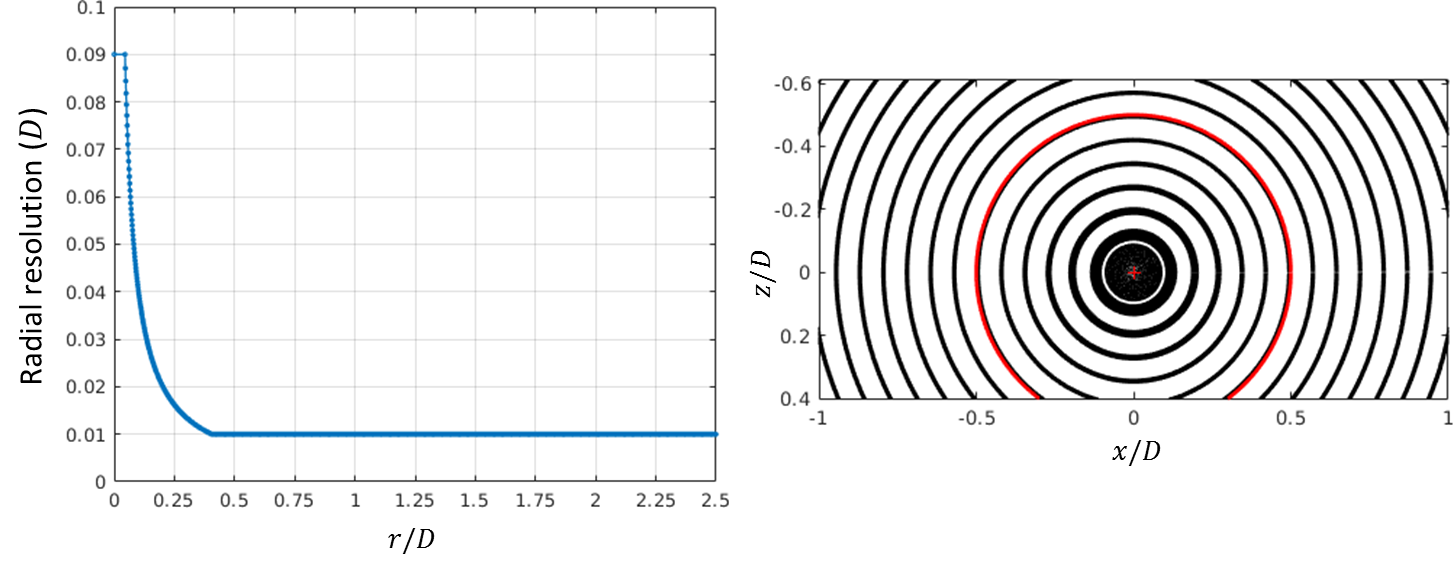} 
		\caption{Geometrical definition of bins used for statistical averaging of the DF-TPTV results. Left: bin radial resolution, $BS_r(r)$ (expressed in jet diameter units), as a function of $r/D$. Right: Layout of in a cross-sectional plane. Bins, which have been subsampled for clarity, are depicted as black rings whose thickness is equal to their radial resolution $BS_r(r)$. The red circle denotes the jet nozzle.}
	\label{fig:BinDef}
\end{center}
\end{figure*}
\begin{figure*}[htbp]
\begin{center}\leavevmode
\includegraphics[width=0.9\textwidth]{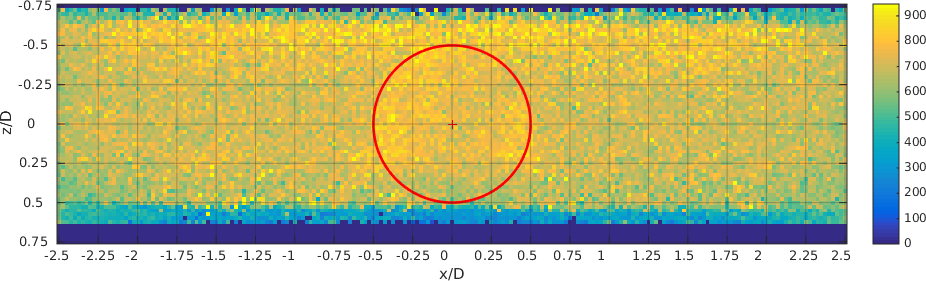} 
		\caption{Intensity repartition of reconstructed particles in cross-sectional plane $y/D=1.4$, averaged over square bins of $0.025D\times 0.025D$ size. The red circle denotes the jet nozzle.}
	\label{fig:LightIntensity}
\end{center}
\end{figure*}

For DF-TPTV, which yields scattered vector data, we resort to bin averaging, as traditionally done in PTV methods \citep{Kahler16,Aguera16,Jux18}. In the present jet flow context, we choose a specific form of bins in order to increase the number of samples. We exploit the assumption of statistical axisymmetry to define them as annuli of increasing radius in a cross-sectional plane. More precisely, introducing the radius $r=\sqrt{x^2+z^2}$, a bin centered at radius $r$, with radial resolution  $BS_r(r)$ and streamwise resolution $BS_y$, is defined as the volume:
  \beq \nonumber
  \left[r - \frac{BS_r(r)}{2};\,r + \frac{BS_r(r)}{2}  \right]
  \times \left[\frac{y}{D} - \frac{BS_y}{2};\,\frac{y}{D}  + \frac{BS_y}{2}  \right]
    \eeq 
Another specificity is that, as depicted in figure \ref{fig:BinDef} (right), these bins are not all strictly annuli, but rather truncated annuli for the majority of them, as vectors outside of the range $-0.61\leq z\leq 0.40$ have been excluded. This choice is justified by the repartition of light intensity in the volume, which was slightly asymmetrical with respect to the jet centre. As figure \ref{fig:LightIntensity} shows, intensity drops quite rapidly for $z/D>0.4$, which leads to spurious end effects in the motion estimation there within DF-TPTV. This was observed in turn to yield less reliable results, in spite of the rejection (which was also observed to exclude more samples); indeed, values of the fluctuating velocity were observed to be spuriously higher when including this region in the estimation.

In order to gather enough samples in each bin for convergence, while introducing as little spatial smoothing as possible, we choose $BS_r(r)$ to decrease from $0.09D$ for $r/D=0$, to $0.01D$ for $r/D\geq{0.4}$, as seen in figure \ref{fig:BinDef} (left). Note that, although $BS_r$ decreases with $r$ in the jet core and then stabilizes in the shear layer and the outer flow, bins are nearly of same volumic extent in the whole field, as a result of the radial geometry. Value $BS_r(r>0.4)=0.1$ has been observed to be the smallest reachable, i.e. preserving enough samples for a satisfactory level of convergence. Also, we observed that the value of maximum RMS velocity in the shear layer is nearly insensitive to moderate variations of $BS_r$ around this value. Also, in order to allow fine sampling within the shear layer, we consider a high overlap between the bins, equal to at least $85 \% $. The streamwise resolution $BS_y$ is set to $0.084D$, which corresponds to the interrogation window (IW) size of the planar PIV processing (see below).

For each bin, we compute the mean and variance of each velocity component as (here exemplified on the streamwise component): 
\begin{equation}
      \overline{v'^2} = \frac{1}{M-1} \sum_{t=0}^N \sum_{i=1}^{N_t} ({v}(t)_i - V)^2
  \end{equation}
  where 
\begin{equation}
     V  = \frac{1}{M} \sum_{t=0}^N \sum_{i=0}^{N_t} {v}(t)_i \, \mbox{and}\, M = \sum_{t=0}^N {N_t}
  \end{equation}
where ${v}(t)_i$ denotes the velocity of the $i^{th}$ vector measured in the considered bin at time $t$, $N_t$ is the total number of vectors in the bin at time $t$ and $N$ the number of snapshots.

For consistency, images acquired in the run for planar PIV are processed with rectangular, top-hat IWs of $31\times{5}$ pixels ($0.084D\times 0.014D$, respectively in the $y$ and $z$ directions). With such a parameter, the IW size is as close as possible to the bin size of DF-TPTV, although resolution in $z$ is slightly higher. For this planar PIV run, $3000$ snapshots were acquired, in order to reach full convergence of mean and fluctuating velocities.

\subsubsection{Compared velocity profiles in the jet near field}

To compare mean and RMS velocities yielded by DF-TPTV and planar PIV, we restrict to the jet near field, here $y/D\leq{2}$. As a matter of fact, in this zone the turbulence rate remains moderate and can be estimated with a satisfactory accuracy with DF-TPTV, which is characterized by the lowest number of samples per bin.
\begin{table*}[h]
\centering
\begin{tabular}{cc}
\includegraphics[width=0.9\columnwidth]{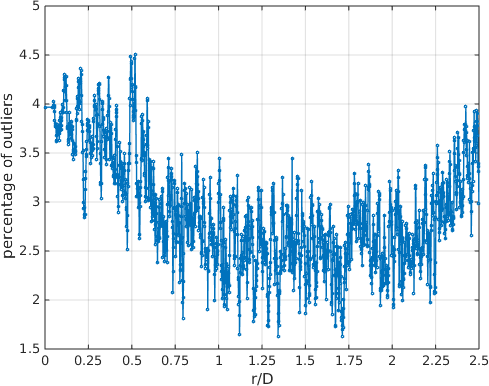} &\includegraphics[width=0.9\columnwidth]{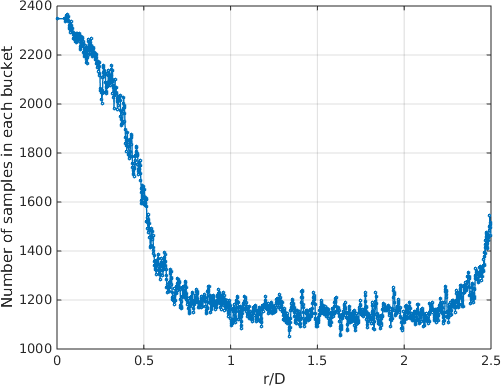}
\end{tabular}
		\captionof{figure}{Percentage of rejected vectors per bin (left), and number of vectors per bin after rejection (right) as a function of $r/D$, in cross-sectional plane $y/D=1.4$, yielded by the DF-TPTV method.}
	\label{fig:BinsSampleRejet}
\end{table*}

Figure \ref{fig:BinsSampleRejet} shows the number of samples obtained by DF-TPTV in each bin, as well as the percentage of vectors rejected, as a function of $r/D$, at $y/D = 1.40$. As a result of the difference in average seeding density between the jet and the outer flow (see also section \ref{sec:expesetup}), up to maximum $2350$ samples per bin are obtained in the jet core, progressively decreasing in the outer shear layer and the ambient flow, here to $1100-1200$. Rejection also has a slightly different behaviour in the jet core and in the external flow. Around $3.5$ to $4\%$ of the initial vectors are discarded in the core, while the average rejection rate in the outer flow is rather of the order of $2.5\%$, with a slight increase in the outermost region. This is again consistent with the difference in seeding between these two zones, which translates immediately in an increase in the average measurement error (see in particular figure \ref{fig:velerror}), therefore inducing more rejection where seeding is denser. Similar trends and orders of magnitude are observed at all other stations in $y/D$ shown hereafter.

\begin{table*}[h]
\begin{tabular}{cc}
\includegraphics[width=\columnwidth]{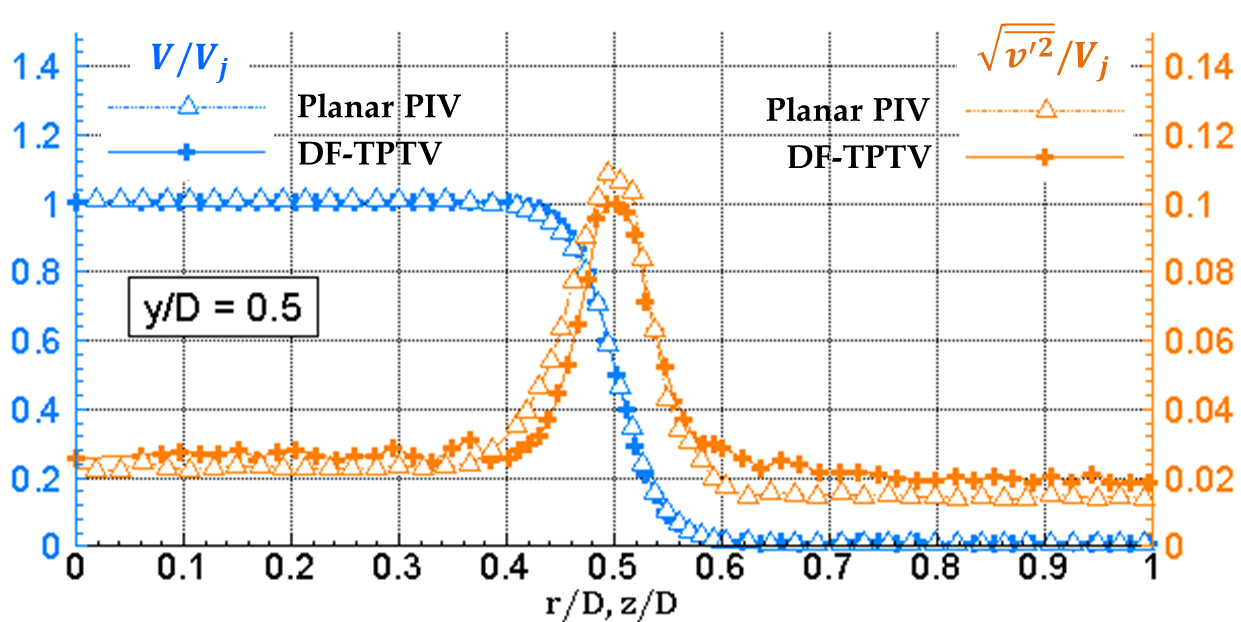} & \includegraphics[width=\columnwidth]{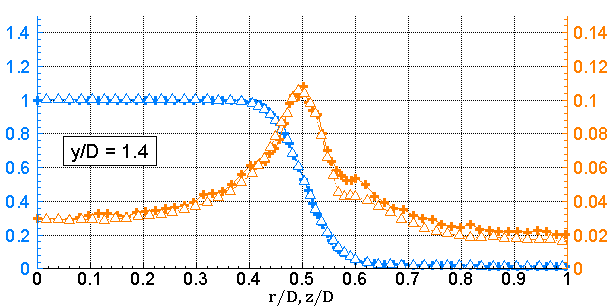}       \\
\includegraphics[width=\columnwidth]{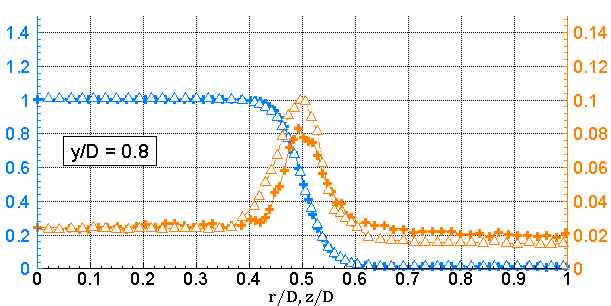} & \includegraphics[width=\columnwidth]{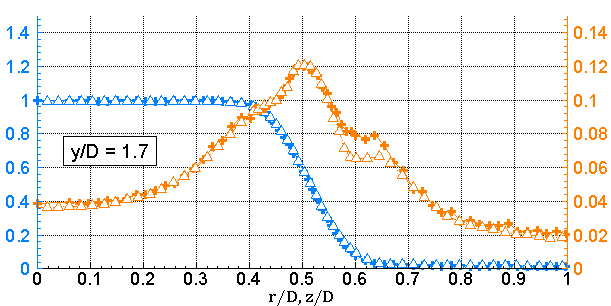} \\
\includegraphics[width=\columnwidth]{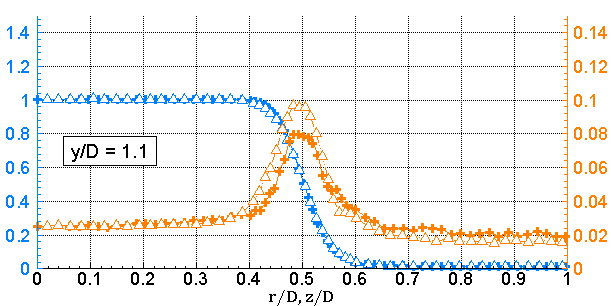} & \includegraphics[width=\columnwidth]{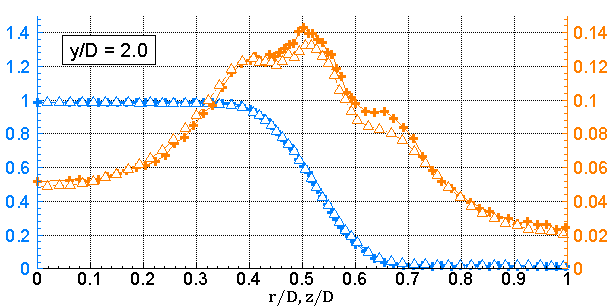}
\end{tabular}
\captionof{figure}{Streamwise mean and fluctuating velocity profiles, along a jet radius, for various cross-sectional locations $y/D$ in the near-field. Comparison between DF-TPTV, which uses bin averaging with statistical axisymmetry assumption ($r/D$ abscissa), together with planar PIV ($z/D$ direction). To ease readability and comparison, a horizontal coordinate shift has been applied, see the text for more details.}
\label{fig:CompDFTPTV_PIV2D}
\end{table*}

Figure \ref{fig:CompDFTPTV_PIV2D} shows profiles of the mean and fluctuating streamwise velocity in the radial (resp. $z$) direction, yielded by DF-TPTV (resp. planar PIV). As the two measurements have been performed with different calibrations, slight discrepancies in their respective frame of reference have been detected. To compensate for these and ease comparison, a shift in abscissa has been performed on the curves. Its value has been determined so that the value of the mean velocity at the centre of the shear layer matches for both measurements. In other words, denoting $r_{1/2}$ (resp. $z_{1/2}$) the position such that $V$ is equal to $V(0)/2$, this shift has been chosen such that $r_{1/2}=z_{1/2}$. Consistently with the choice for bin or IW size, it is observed that the mean velocity profiles match almost perfectly. Only very minor discrepancies appear, with the profiles of planar PIV characterized by slightly larger shear layers, this being in line with the slightly larger horizontal resolution (see table \ref{tab:AveragingResSamples}).

Regarding the fluctuating streamwise velocity or turbulence rate, curves for DF-TPTV and planar PIV are observed either to collapse, or to differ by up to roughly $0.02$. These maximum discrepancies are observed in a localized zone in the outer shear layer ($r/D,z/D\approx 0.6-0.7$ for $y/D=1.7$, and with a lower magnitude at $y/D=1.4$ and $2$), and also within the shear layer for $y/D=0.8$ and $1.1$. It should be noted that they cannot be ascribed to the partially converged character of $\sqrt{\overline{v'^2}}/V_j$ estimated by DF-TPTV. Indeed, monitoring of its convergence with respect to the number of samples yields an estimate of the statistical uncertainty of roughly $0.001-0.002$. Turning now to the detailed explanation of these curves, a first, most directly understandable observation is the level of turbulence rate in the jet core ($r/D,z/D<0.3-0.4$) and in the outer flow ($r/D,z/D>0.8$) for the most upstream locations, say up to $y/D=1.1$. Indeed, flow in these zones should be strictly laminar and therefore $\sqrt{\overline{v'^2}}/V_j$ should vanish. Levels observed there are thus measurement noise and can be directly compared, in the case of DF-TPTV, to the levels of the RMS error on velocity from synthetic data, $RMS_v$, presented in section \ref{sec:accuracy} (see figure \ref{fig:velerror}). One observes, in the present experiment, values close to $0.025$ in the core, and $0.02$ in the outer flow, which correspond to $0.13$ and $0.1$ voxel, respectively. Firstly, the higher value in the jet core is again consistent with the denser average seeding there. Also, both these values compare well with the estimation of $RMS_v$ on the true particles only: this quantity indeed ranges from $0.11$ voxel for ppp $=0.03$, to $0.20$ voxel for ppp = $0.06$, which is the range of image seeding densities here. The fact that the present experimental noise level are closest to the lowest theoretical value for $RMS_v$ (or even slightly lower than it) might have two origins. Firstly, during the run, seeding density is found to decrease quite rapidly, so that the average density is closer to the final value of $0.03$ ppp than to the initial $0.06$ ppp. Moreover, the slightly lower value of noise compared to $RMS_v$ found in the outer flow is probably due to the fact that the image pre-processing performed before DF-TPTV (see section \ref{sec:acqprocpar}) should lead to a lowering of the effective seeding density. Note that these values also confirm the very good efficiency of the outlier rejection procedure, as the present comparison is performed here with $RMS_v$ values estimated by only keeping the true vectors, and, overall, illustrate the excellent robustness of the DF-TPTV method in this experimental context. Finally, it is interesting to observe that DF-TPTV exhibits a noise level which is very similar to that of planar PIV processed with FOLKI-PIV, as long as the bin size and the IW size are taken equal. In the curves, these noise levels are seen to be nearly equal, or slightly higher for DF-TPTV (note that the IW size for planar PIV is in fact very slightly larger than the bin size, as shown in table \ref{tab:AveragingResSamples}). This fact also helps to understand why the agreement between results yielded by the two methods is very close in a general way. Possible reasons for zones with discrepancies could be a partial lack of axial symmetry of the jet, making the estimation by bin averaging of DF-TPTV less accurate, or a different sensitivity of DF-TPTV and of planar PIV to flow gradients.

\section{Conclusion}\label{sec:conclusion}

We proposed here a novel technique for performing 3D PTV from traditional double frame images, termed double frame tomographic-PTV (DF-TPTV). Its main specificity is that it takes advantage of the sparse nature of the tomographic PIV problem. It first produces spiky particle reconstructions located on a voxel grid, leveraging on the PVR model \citep{Champagnat14} and the sparsity-based algorithm LocM-CoSaMP \citep{Cornic15}. Reconstructed particles are matched using a low resolution predictor yielded by 3D correlation before being accurately localized through a global optimization procedure. Good performances have been obtained over a large range of seeding densities (ppp $\in (0, 0.08)$) on synthetic images generated using a DNS data of a turbulent channel flow. DF-TPTV has been then demonstrated to operate successfully at ppp as high as $0.06$ on experimental data on a round air jet ($Re=4600$). In this experiment, the DF-TPTV technique allowed to produce both reliable instantaneous velocity vector fields and accurate ensemble statistics, upon introducing an additional outlier rejection step. Statistical results have been obtained by a specific bin averaging process exploiting the jet average axisymmetry, and were found in excellent agreement with reference measurements by planar PIV at comparable spatial resolution.


Overall, we would like to emphasize that DF-TPTV is a particle tracking technique that relies on the same amount of information as TomoPIV, not only in terms of hardware (double frame acquisition, either at low or high frequency), but also of seeding densities, since it yields reliable results for ppp up to $0.06-0.08$. A consequence is that, within a given experiment, there is no need to perform separate runs at lower densities in order to perform DF-TPTV as well as TomoPIV processing.

A possible direct perspective to this work could be to further improve of the method's robustness to low signal-to-noise ratios, so as to exploit the maximum from a given volumic illumination, often less intense on the volume edges, as was the case in the present experiment. Future research paths will also target a more drastic increase in the capability of DF-TPTV to characterize complex turbulent flows, on an instantaneous point of view. This will be tackled by proposing new data assimilation schemes, performing physically-sound instantaneous interpolation between the obtained scattered vectors. 


\bibliographystyle{spbasic}      
\bibliography{Bibliography}
\end{document}